\documentstyle[a4,11pt,epsf]{article}
\newcommand{\be}{\begin{eqnarray}}
\newcommand{\ee}{\end{eqnarray}}
\newcommand{\nn}{\nonumber}

{\newcommand{\lsim}{\mbox{\raisebox{-.6ex}{~$\stackrel{<}{\sim}$~}}}
{\newcommand{\gsim}{\mbox{\raisebox{-.6ex}{~$\stackrel{>}{\sim}$~}}}

\def\0n{0\nu\beta\beta}

\begin{document}
\begin{center}
{\Large \bf Neutrino Majorana Mass and Baryon Number of the Universe 
below the Electroweak Symmetry breaking Scale} \\
\vspace{0.5cm}
{\large H.V. Klapdor-Kleingrothaus $^{\dagger}$, St. Kolb $^{\dagger}$ 
and U. Sarkar$^{\dagger,\ddagger}$}\\
\vspace{0.3cm}
{$^{\dagger}$ Max-Planck-Institut f\"ur Kernphysik,
P.O. 10 39 80, D-69029 Heidelberg, Germany} \\ \vspace{0.3cm}
{$^{\ddagger}$ Physical Research Laboratory, Ahmedabad, 380 009 India}
\end{center}

\vspace{0.5cm}

\begin{abstract}

If the neutrino is Majorana type and the electroweak phase transition
is second or weak first order, neutrino-induced interactions together
with sphaleron transitions have the potential to erase a previously 
generated baryon asymmetry of the universe. Taking
correctly into account the evolution of the vacuum expectation of the
Higgs field 
%any entry of the left-handed Majorana neutrino mass matrix 
%is constrained to be less than of order ${\cal O}(10 MeV)$. 
the effective light neutrino masses are constrained to be
lighter than ${\cal O}(10 MeV)$, while the effective heavy masses 
are constrained to be heavier than ${\cal O}(10^7 GeV)$.
\end{abstract}

\vspace{1cm}

Recent experiments seem to indicate that the neutrino is massive
\cite{kamiokande}. From a model-building point of view the most 
natural structure of the neutrino mass matrix contains both 
Lepton-number- ($L$-) conserving Dirac- and 
$L$-violating Majorana-type entries (see {\it e.g.} \cite{moha}). 
In principle observable consequences are $L$-violating processes 
such as neutrinoless double beta ($\0n$) decay and $L$-violating
lepton-gauge boson scattering (inverse $\0n$ decay).

On the other hand, in the early universe $L$ violation together 
with sphaleron-mediated transitions has the potential to create 
the Baryon number ($B$) of the universe (BAU) (Leptogenesis)
or erase an existing BAU. The latter may be the case both above
or below the electroweak symmetry breaking scale. This
consideration would limit the amount of $L$ violation
and hence give a bound on the Majorana mass of the neutrinos. In the
following we will reconsider the limits on neutrino masses and 
point out that the evolution of the vacuum expectation value 
($vev$) of the Higgs responsible for electroweak symmetry breaking
weakens the existing bound by as large as three orders of magnitude.

In an extension of the standard model, the Majorana mass of
the neutrinos can come from an effective dimension-5 operator
\cite{dimfive} 
\be \label{dimfive}
\frac{\alpha_{ij}}{M}(L^T_i C^{-1} \tau_2 \vec{\tau} L_j)
                     (H^T \tau_2 \vec{\tau} H)
\ee
where $i,j = 1,2,3$ are generation indices,
$M$ is the $L$-violating mass-scale, $\alpha$ is an effective
coupling and $L$ ($H$) are $SU(2)_L$ lepton (Higgs) doublets.
After the electroweak symmetry breaking, when the higgs doublet
scalar acquires a $vev$, the neutrinos get a Majorana mass of
the order of 
\be 
m_\nu \sim \frac{\alpha_{ij} \langle v(T) \rangle^2}{M} 
\ee
where $\langle v(T) \rangle$ is the $vev$ of $H$.
The condition that the associated $L$ violation should 
not wash out the primordial BAU then gives an upper bound on the Majorana
mass of the neutrinos of the order of a few keV \cite{utpal,lgb}.
%Most of these analysis simplified the analysis by taking the $vev$
In most of these approaches the analysis was simplified by assuming the
$vev$ to be constant and the rate of the $L$-violating interactions
was considered to be less than the expansion rate of the universe.
However, above the critical temperature $T_C$ of the electroweak
phase transition (EWPT) the $vev$ of $H$ is zero. Below $T_C$ the $vev$ 
starts growing. On the other hand soon the sphalerons freeze out and they 
cannot wash out the BAU any longer. Thus, during the period when the 
sphalerons wash out the BAU, the $vev$ may still be quite small. This 
weakens the upper bound on the Majorana mass of the neutrinos. In the
following this argument will be discussed in more detail.

%{\tt St.K. probably simply skip this statement or make it compatible
%with what follows:}
%The Majorana part of the mass matrix coupling $SU(2)_L$ doublet 
%neutrinos is effective dimension five and reads above the electroweak
%symmetry breaking scale (EWS) \cite{dimfive} 
%($i$,$j$ generation indices)

Consider the see-saw mechanism of neutrino masses \cite{seesaw}.
The $L$-violating mass scale will have its origin 
from integrating out heavy $SU(2)_L$ singlet neutrinos.
The neutrino mass-matrix has the general structure
\be \label{structure}
{\cal M}^{\nu}=\left( \begin{array}{cc} 0 & (m^D)^T \nn \\
                                        m^D & M \end{array} \right)
\ee
where $m^D \sim \langle v(T) \rangle$ is the Dirac mass matrix coupling 
$SU(2)_L$ doublet neutrinos $\nu$ to $SU(2)_L$ singlet neutrinos $N$, 
whereas $M$ is a Majorana mass matrix for the $N$'s. 
Below the EWS the Higgs field aquires a (temperature dependent)
vacuum expectation value $\langle v (T) \rangle$ and 
%operator 
%(\ref{dimfive}) results in a Majorana mass matrix 
%$m^M_{ij} = \alpha_{ij} \langle v (T) \rangle^2/M$ for 
%the $SU(2)_L$ doublet neutrino fields. If $M$ pertains to 
%the $SU(2)_L$ singlet neutrinos this is just the see-saw 
%mechanism. 
so that $\nu$ and $N$ mix with resulting masses
\be \label{states}
m_i = \sum_j U_{ij} {\cal M}_{ij}\ , \ U U^{\dagger} = 1
\ee
If the scale of $M$ is much larger than that of $m^D$ the 
diagonal mass-matrix consists of two blocks of light and heavy
masses (see-saw mechanism)
\be
m_{light} \approx - (m^D)^T M^{-1} m^D \ , \ m_{heavy} \approx M \ .
\ee
The off-diagonal blocks of the mixing matrix $U$ are approximately
$(m^D)^{\dagger}(M^{-1})^{\dagger}$ and $-M^{-1}m^D$. 

In the triplet
higgs model \cite{trip} one introduces a triplet higgs scalar $\xi$ with
mass $M$. The couplings of the triplet higgs breaks lepton number
explicitly at the scale $M$, but the $vev$ of the triplet higgs
gets a see-saw contribution of amount
\be
\langle \xi \rangle \sim {\langle H \rangle^2 \over M}  .
\ee
The direct coupling of the triplet higgs with the two neutrinos
then give a Majorana mass to the neutrinos. 

%In a model without $SU(2)_L$ singlet neutrinos masses may be 
%generated via an effective dimension five operator ({\tt St.K. see
%above, make this consistent}) 
%\be \label{dimfive}
%\frac{\alpha_{ij}}{M}(L^T_i C \tau_2 \vec{\tau} L_j)
%                     (H^T \tau_2 \vec{\tau} H)
%\ee
%where $M$ now is a heavy mass-scale tied to the origin of operator
%(\ref{dimfive}), $\alpha$ is an effective coupling and $L$ ($H$) 
%are $SU(2)_L$ lepton (Higgs) doublets. As in the previous case below
%the EWS the neutrino mass is proportional to the square of the Higgs
%$vev$.

In the early universe neutrinos give rise to $L$-violating processes
such as ($i$,$j$ are generation indices) 
\be \label{process}
e_i^{\pm} e_j^{\pm} \leftrightarrow W^{\pm} W^{\pm} \ .
\ee
The masses $m_{ij}^M$ give rise to $L$-violating processes such as
neutrinoless double beta decay ($\0n$) (for an overview see 
{\it e.g.} \cite{klap}).
For a linear collider this process has been studied in \cite{inverse}.   
%$e^{\pm} e^{\pm} \leftrightarrow W^{\pm} W^{\pm}$ \cite{inverse}. 

In the early universe $m_{ij}^M$ induced processes have the potential
to erase the obeserved asymmetry in the baryon- ($B$) and 
antibaryon-number of the
universe (BAU) (see {\it e.g.} \cite{kolb}). This is due to the fact
that as long as sphaleron transitions are in thermal equilibrium
\cite{sph} $B$ and $L$ are both proportional to $(B-L)$ \cite{ht}. Hence, if
$L$ is erased by an $L$-violating process and sphalerons are still
operative, $B$ is erased as well. If the electroweak phase transition
(EWPT) with an associated critical temperature $T_C$ is strong first order 
sphalerons are never in thermal equilibrium and $L$-violating processes
do not affect $B$ below $T_C$. On the other hand, if the EWPT is second
or weak first order there is a period 
\be \label{range}
T_{out} < T < T_C
\ee
($T_{out}$ denotes the sphaleron freezing-out temperature) during which 
$L$-violating processes have the potential to erase $L$ and consequently 
$B$. Hence, if the EWPT is second or weak first order the requirement
that a preexisting BAU should not be washed out poses a limit on the
amount of $L$ violation. In the case of a Majorana neutrino in previous
works \cite{utpal} an estimated limit $m_{ij}^M \lsim 20 keV$ has been 
obtained. In this note it will be argued that this bound is in fact 
three orders of magnitude less stringent if the temperature dependence
of the $vev$ is correctly taken into account as has been done recently
for the case of $L$-violating sneutrinos in \cite{barpap}. 

\begin{figure}[t]
\vspace*{-1cm}
\hspace*{3cm}
\epsfxsize65mm
\epsfbox{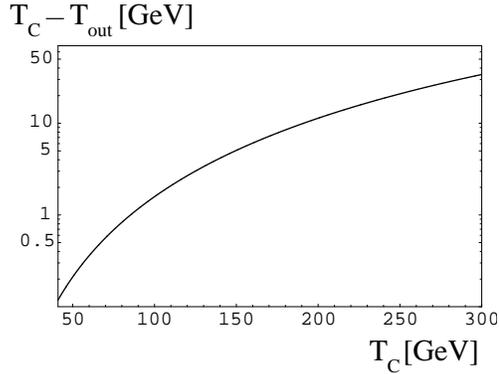}
\vspace{-0.5cm}
\caption{The difference of the critical temperature $T_C$ and 
the sphaleron 
freezing out temperature $T_{out}$ determined by 
$\Gamma_{sph}(T_{out})=H(T_{out})$ in dependence of $T_C$.}
\label{rangeplot}
\end{figure}

The temperature dependence of the $vev$ for a second or weak first
order phase transition is given by
\be \label{vacvalue}
\langle v (T) \rangle \approx \langle v (T=0) \rangle 
(1 - T^2/T_C^2)^{1/2} \ \ , \ \ \langle v (T=0) \rangle = 246 GeV \ 
\ee
and the the sphaleron rate in the broken phase is \cite{sphrate}
\be \label{sphrate}
\Gamma_{Sph} (T) \approx 2.8 \cdot 10^5 \; T^4 \; \kappa \; 
            \left(\frac{\alpha_W}{4 \pi}\right)^4
            \left(\frac{2 m_W (T)}{\alpha_W T}\right)^7
            \exp\left(-\frac{E_{sp}(T)}{T}\right)
\ee
where 
\be
m_W (T) = \frac{1}{2} g_2 \langle v (T) \rangle \ ,
\ee
the free energy of the sphaleron configuration is given by
\be
E_{Sph} (T) = \frac{2 m_W (T)}{\alpha_W} B \left(\frac{m_H}{m_W}\right),
\ee
$B(0)=1.52,\; B(\infty)=2.72$ and $\kappa$=exp(-3.6) \cite{moore}.
As usual $T_{out}$ is determined by the condition
\be \label{sphfreezeout}
\Gamma_{sph} (T_{out}) = H(T_{out}) = 
                         1.7 \sqrt{g_*}\frac{T_{out}^2}{M_{Pl}} 
\ee 
where $M_{Pl} \approx 10^{19}GeV$ is the Planck scale and 
$g_* \approx 100$ in the Standard Model. Lattice simulations suggest 
that for a Higgs mass of around 
$m_H \approx 70 GeV$ $T_C \approx 150 GeV$ and higher for larger 
values of $m_H$ \cite{valuetc}. For our phenomenological purposes   
$T_C$ will be varied between $50 GeV$ and $250 GeV$. The temperature
range eq. (\ref{range}) is plotted in figure \ref{rangeplot}. It
is smaller than $1GeV$ for $T_C \lsim 100 GeV$ but of order
${\cal O}(10 GeV)$ for $T_C \gsim 200 GeV$.  

Relevant processes for depleting a pre-existing $L$ number during the
epoch (\ref{range}) are $L$-violating $2 \leftrightarrow 2$ scatterings 
$W^{\pm} W^{\pm} \leftrightarrow e^{\pm}_i e^{\pm}_j$, 
$W^{\pm} e_i^{\mp} \leftrightarrow W^{\mp} e_j^{\pm}$ and gauge
boson decays. The depletion of an initial 
$L$ number 
$L_i$ is described by (see for example \cite{kolb}) 
\be \label{evolution}
L(z)=L_i \exp \left[ 
      - \int\limits_{z_c}^{z_{out}} d z' z' 
                 [g_* \frac{n_{\tilde{\nu}}}{s} % \gamma_D 
                  \Gamma_D(z')+ n_{\gamma} 
                  \langle \sigma |v| \rangle)]/H(T=m_{\tilde{\nu}}) \right]
\ee
Working with the Boltzmann equation will then give us the amount of
residual asymmetry after the spheleron transitions have frozen out. 
Unlike earlier works, where the interaction rate has been compared with 
the expansion rate of the universe, we consider the condition for 
erasure of the primordial BAU is that the asymmetry depletes by a
factor of at least 10. In most of the cases when we get the bound,
the depletion is more than two orders of magnitude.

For the see-saw mechanism case the thermally averaged contribution of 
the light neutrino states with masses $m_k \ll T_C$ may be approximated by
\be \label{contributionlight}
\langle \sigma |v| \rangle_{ij} \approx 
       \frac{\alpha_W^2 \langle m_{ij} \rangle ^2}{T^4} \ , \
\langle m_{ij} \rangle = \sum_k U_{ik} U_{jk} m_k 
\ee
and the contribution of the heavy states with masses $M_n \gg T_C$ is
\be \label{heavycontribution}
\langle \sigma |v| \rangle \approx 
       \frac{\alpha_W^2}{\langle M_{mn} \rangle ^2} \ , \
\frac{1}{\langle M_{ij} \rangle} = \frac{\sum_n U_{in} U_{jn}}{M_n}
\ee

\begin{figure}[t]
\epsfxsize140mm
\epsfbox{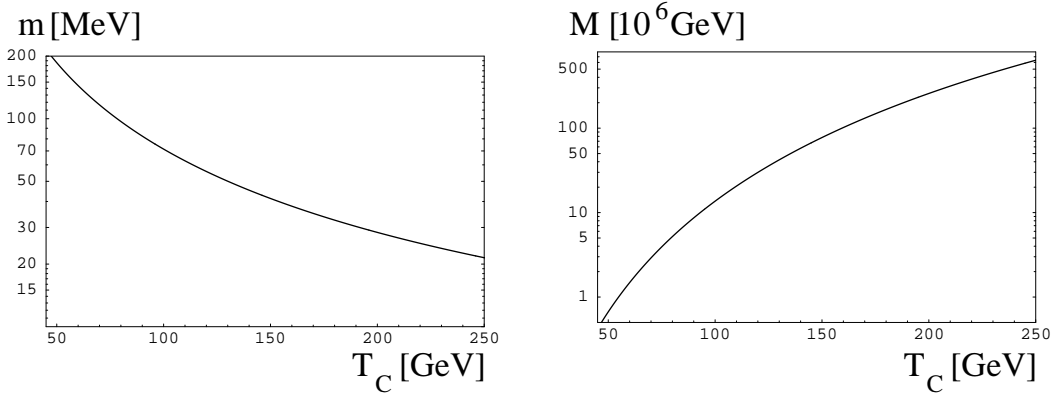}
\vspace{-0.5cm}
\caption{The bounds on the superposition of light neutrino states 
$\langle m_{ij} \rangle$ (left, region above the curve excluded)
and of heavy neutrino states $\langle M_{ij} \rangle$ (right, region
below the curve excluded)}
\label{majneut}
\end{figure} 

\noindent
where $\alpha_W$ is the weak coupling constant. Compared to the
zero-temperature case both cross-sections are suppressed by a
factor 
\be \label{suppression}
\frac{\langle v(T) \rangle ^2}{\langle v(T=0) \rangle ^2}=
\Big( 1-\frac{T^2}{T_C^2} \Big)^2 \ ,
\ee
since for the light states 
$m_k \sim \langle v(T) \rangle ^2$ and for the heavy states
$U_{in} U_{jn} \sim \langle v(T) \rangle^2$, see above. The bounds
on the quantities $\langle m_k \rangle$ and $\langle M_n \rangle$
are displayed in figure \ref{majneut}. For plausible values of the
critical temperature $T_C \gsim 150 GeV$ the bound on the light
states is of order $\langle m_{ij} \rangle \lsim {\cal O} (10 MeV)$
while the bound on the heavy states is of order 
$\langle M_{ij} \rangle \gsim 10^7 GeV$. 

For the triplet higgs mechanism, the bound on the mass of the
neutrino comes out to be the same as the bound on the light 
neutrino state of the see-saw mechanism. In this case also the
suppression is given by equation (\ref{suppression}), since the
neutrino mass is again proportional to $m_\nu \propto \langle
v(T) \rangle^2$. Thus even in this case the bound comes out to
be around 10 MeV. Given the generality of the dimension-5 
operators for the Majorana neutrino mass, one may conclude 
that in all models of neutrino masses this bound is valid.
Previous estimates of the
light neutrino masses have been given in \cite{utpal} for the case
eq. (\ref{dimfive}) and it has been argued that for every entry of
the corresponding mass matrix the bound from the BAU is of order
$m_{ij} \lsim 10 keV$, that is three orders of magnitude more stringent
than if the evolution of the $vev$ is taken into account. 

In summary, we included the effect of evolution of the higgs
$vev$ and solved the Boltzmann equation in estimating the bound 
on the neutrino masses coming from the erasure of the baryon asymmetry
of the universe. This makes the bounds three orders of magnitude 
weaker than the one obtained from earlier naive estimates.\\

\centerline{\bf Acknowledgement}

\vspace{0.5cm}
%\noindent
U.S. wants to thank Max-Planck-Institut f\"ur Kernphysik for hospitality.


\begin{thebibliography}{99}
\bibitem{kamiokande} Y. Fukuda {\it et al.}, the Super-Kamiokande
                     Collaboration, Phys. Lett. {\bf B 467} (1999) 185.

\bibitem{moha} R.N. Mohapatra and P.B. Pal, ``Massive Neutrinos
               in Physics and Astrophysics'' (World Scientific,
               Singapore, 1991).

\bibitem{dimfive} S. Weinberg, Phys. Rev. Lett. {\bf 43} (1979) 1566;
                  E. Akhmedov, Z. Berezhiani and G. Senjanovi\v{c}, 
                  Phys. Rev. Lett. {\bf 69} (1992) 3013;
                  E. Ma, Phys. Rev. Lett. {\bf 43} (1998) 1171.

\bibitem{utpal} M. Fukugita and T. Yanagida, Phys. Rev. {\bf D 42} (1990) 1285;
S.M. Barr and A.E. Nelson,  Phys. Lett. {\bf B 246} (1991) 141;
W. Fischler, G. Giudice, R. Leigh and
S.  Paban, Phys. Lett. {\bf B 258} (1991) 45; W.  Buchm\"{u}ller
T. and Yanagida, Phys. Lett. {\bf B 302} (1993) 240;
U. Sarkar, Phys. Lett. {\bf B 390} (1997) 97.

\bibitem{lgb} B. Campbell, S. Davidson, J. E. Ellis and K. Olive,
   Phys. Lett. {\bf B 256} (1991) 457;
  E. Ma, M. Raidal and U. Sarkar  Phys. Lett.  {\bf B 460} (1999) 359
 H. Dreiner and G.G. Ross, Nucl. Phys. {\bf B 410} (1993) 188;
 J.M. Cline, K. Kainulainen and K.A. Olive Phys. Rev.  {\bf D 49} (1994) 6394;
   A. Ilakovac and A. Pilaftsis, Nucl. Phys. {\bf B 437} (1995) 491.

\bibitem{seesaw} M. Gell-Mann, P. Ramond and R. Slansky, in {\it
  Supergravity},  Proceedings  of the Workshop,  Stony Brook, New
  York, 1979,  ed. by P.  van  Nieuwenhuizen  and D.  Freedman
  (North-Holland, Amsterdam);  T. Yanagida, in {\it Proc of
  the  Workshop  on Unified  Theories  and  Baryon  Number in the
  Universe},  Tsukuba,  Japan, 1979, edited by A.  Sawada and A.
  Sugamoto (KEK Report No.  79-18, Tsukuba);
   R.N. Mohapatra and G. Senjanovi\'{c}, Phys. Rev. Lett. {\bf 44} (1980) 912.

\bibitem{trip} E. Ma and U. Sarkar, Phys. Rev. Lett. {\bf 80} (1998) 5716; 
               G. Lazarides and Q. Shafi, Phys. Rev. {\bf D 58} (1998) 
               071702.

\bibitem{klap} H.V. Klapdor-Kleingrothaus in Proc. Int. Conf. on
               Lepton- and Baryon-Number Non-Conservation, Trento, 
               Italy, April 20-25, 1998, IOP, Bristol (1999);
               H.V. Klapdor-Kleingrothaus, Int. J. Mod. Phys.
               {\bf A 13} (1998) 3953. 

\bibitem{inverse} G. Belanger, F. Boudjema, D. London and 
                  H. Nadeau, Phys. Rev. {\bf D 53} (1996) 6292;
                  C. Greub and P. Minkowski, Int. J. Mod. Phys. {\bf A 13} 
                  (1998) 2363.

\bibitem{kolb} E.W. Kolb and M.S. Turner, ``The Early Universe''
               (Addison-Wesley, Redwood City, CA, 1990).

\bibitem{ht} S. Yu. Khlebnikov and M.E. Shaposhnikov, Nucl. Phys. 
             {\bf B 308} (1988) 885;
             J.A. Harvey and M.S. Turner, Phys. Rev. {\bf D 42} (1990) 
             3344.

\bibitem{sph} V.A. Kuzmin, V.A. Rubakov and M. Shaposhnikov, 
              Phys. Lett. {\bf B 155} (1985) 36.

\bibitem{barpap} H.V. Klapdor-Kleingrothaus, St. Kolb and V.A.
                 Kuzmin, Phys. Rev. {\bf D}, in press and hep-ph/9909546.

\bibitem{sphrate} V.A. Rubakov and M.E. Shaposhnikov,
                  Usp. Fiz. Nauk {\bf 166} (1996) 493, 
                  Phys. Usp. {\bf 39} (1996) 461; 
                  A. Riotto, hep-ph/9807454.

\bibitem{moore} G.D. Moore, Phys. Rev. {\bf D 59} (1999) 014503.

\bibitem{valuetc} K. Kajantie, M. Laine, K. Rummukainen and 
                  M. Shaposhnikov, Nucl. Phys. {\bf B 466} (1996) 189.


\end{thebibliography}
\end{document}